# Controlled formation of multi-scale porosity in ionosilica templated by ionic liquid

Shilpa Sharma,[1] Julian Oberdisse,[2*] Johan G. Alauzun,[1] Philippe Dieudonné-George,[2] Thomas Bizien,[3] Cansu Akkaya,[2] Peter Hesemann,[1] Anne-Caroline Genix[2*]

[1]ICGM, Université de Montpellier, CNRS, ENSCM, 34095 Montpellier, France

[2]Laboratoire Charles Coulomb (L2C), Université de Montpellier, CNRS, 34095 Montpellier, France

[3]SOLEIL Synchrotron, L'Orme des Merisiers, Gif-Sur-Yvette, 91192 Saint-Aubin, France

* Corresponding authors: anne-caroline.genix@umontpellier.fr; julian.oberdisse@umontpellier.fr

**Abstract**

Mesoporous systems are ubiquitous in membrane science and applications due to their high internal surface area and tunable pore size. A new synthesis pathway of hydrolytic ionosilica films with mesopores formed by ionic liquid (IL) templating is proposed and compared to the traditional non-hydrolytic strategy. For both pathways, the multi-scale formation of pores has been studied as a function of IL content, combining results of thermogravimetric analysis (TGA), nitrogen sorption, and small-angle X-ray scattering (SAXS). The combination of TGA and nitrogen sorption provides access to ionosilica and pore volume fractions, with contributions of meso- and macropores. We then elaborate an original and quantitative geometrical model to analyze the SAXS data based on small spheres ($R_s$ = 1 – 2 nm) and cylinders ($L_{cyl}$ = 10 – 20 nm) with radial polydispersity provided by the nitrogen sorption isotherms. As a main result, we found that for a given incorporation of templating IL, both synthesis pathways produce very similar pore geometries, but the better incorporation efficacy of the new hydrolytic films provides a higher mesoporosity. Our combined study provides a coherent view of mesopore geometry, and thereby an optimization pathway of porous ionic membranes in terms of accessible mesoporosity contributing to the specific surface. Possible applications include electrolyte membranes of improved ionic properties, e.g., in fuel cells and batteries, as well as molecular storage.

**Keywords:** template-directed synthesis, mesoporosity, ionosilica, ionic liquid, BET, TGA, SAXS

**Introduction**

Porous and especially mesoporous systems have outstanding properties in terms of the large presence of interfaces for adsorption or chemical reactions, and the existence of nanosized channels for molecular or ionic transport, in addition to macroscopic solidity. [1] These assets make them valuable candidates for applications in membrane science, for energy storage (batteries or fuel cells) [2,3] or gaz adsorption. [4] Functionalization of mesopores may be used to widen the field of applications, as reviewed by Ma *et al*, [5] with special emphasis on evaporation and aqueous/liquid phase assembly, as well as applications. Ionic systems have attracted considerable interest due to the possibilities of electrical conductance mediated by ion transport. In particular, mesoporous nanostructured polysilsesquioxanes have been formed by introducing ionizable sites via an amine or ammonium precursor, without adding standard silica precursors such as tetraethyl (TEOS) or tetramethyl orthosilicate (TMOS). [6] These materials are called ionosilica (IS), a highly modulable organosilica scaffold containing ionic building blocks, [7] which have found applications in separation [8] and drug delivery. [9] It is possible to generate porosity within IS via an *in situ* soft templating process based on the presence of ionic liquid during the molecular condensation forming the material, as pioneered by Dai *et al* to form a silica aerogel using a nonhydrolytic ionothermal sol-gel reaction. [10] Sol-gel templating based on an ionic liquid (IL) as template have been used to generate mesoporous silica of different morphologies. [11-17] Most ionic liquids used in this context are salts based on the 1-alkyl-3-methylimidazoliums cation, combined with different anions. In the case of IL with short alkyl group (e.g., butyl as used in this work), the group of Antonietti evidenced a mesoporous wormlike architecture resulting from the IL self-assembly during the reaction. [14] It was suggested that both the formation of hydrogen bonds between the anion and the silica silanol groups and the π-π stacking interaction of the imidazolium rings stabilize the molecular arrangement. While in terms of the silica framework itself, a fractal structure formed by the aggregation of primary silica nanoparticles has been reported, [18] similar to silica aerogels. [19,20]

Room temperature ionic liquids are salts displaying melting temperatures below 100°C due to their bulky and asymmetric ion pairs. ILs possess unique physical and chemical properties, like high conductivity, electrochemical stability, low flammability and vapor pressure. [21] They have been used for various applications in energy storage, [22,23] catalysis, [24,25] or gas adsorption. [26] Also, ILs display self-organization behavior in bulk caused by Coulomb interactions, hydrogen bonding, and amphiphilicity, which depend on specific anion/cation interactions. [27,28] In imidazolium-based ILs, nanoscale segregation of polar (ionic) and apolar (aliphatic) domains can be observed, in particular with imidazolium ions bearing sufficiently bulky alkyl substituents (typically with more than three carbon atoms). [29] The soft templating effect of such structured ILs can be exploited to generate mesoporous ionosilica architectures. Studies of the resulting IS monoliths have been published over the last years. [30-32] Thermogravimetric analysis (TGA) provides quantitative information on filling efficacy. This technique is based on the degradation of IL at high temperatures, typically around 450°C, whereas IS is transformed into silica at higher temperatures but remains essentially present. Consistently, it was found that the higher the IL/precursor molar ratio, the higher the amount of IL confined in the IS scaffold. [31]

Several methods are used to remove the IL template from the mesoporous materials. [33] Thermal calcination is easy to implement [15,34] but has several drawbacks such as structural shrinkage and modification/elimination of any type of functional group. If the porosity is open and interconnected, it is more advantageous to empty the pores by washing procedures with organic solvents, e.g., acetonitrile [13,14] or ethanol [32]. The elimination of the IL can be monitored quantitatively by infrared spectroscopy, or by TGA and SAXS as done here. Once the pores are empty, nitrogen sorption provides access to the specific surface area evaluated by the Brunauer-Emmett-Teller (BET) method, as well as to the total pore volume, radii distributions for cylindrical pores, and thus average pore diameters. [35,36] The distribution of pore radii is commonly based on the evaluation of progressive adsorption using the theory of Barrett, Joyner, and Halenda (BJH). [37] For a given pore size, multilayer adsorption on the pore surfaces with subsequent capillary condensation takes place at a given relative pressure. In this article, we propose a simple mathematical conversion of the BJH-results usually represented in the form of dV/dR where V is the adsorbed volume into a number density of cylindrical pores of radius R and length $L_{cyl}$. This density is the key quantity entering the quantitative structural analysis by small-angle scattering.

Small-angle scattering of X-rays (SAXS) or of neutrons (SANS) is a method of choice to study the nanostructure of materials. [38,39] For this technique to work, the most important property is that the nanostructures are visible to the radiation, in the sense that the material is characterized by spatial fluctuations of the scattering length density, which encodes the local scattering power. The small-angle scattered intensity then reflects the local geometry, which may be particles or pores of any morphology. For instance, long cylinders display a characteristic 1/q-power law observable in the low-q range, while characteristic sizes (like the sphere or the cylinder radius) result in a so-called Guinier law, i.e., intensity plateaus of typical cut-off in q at around the inverse of the radius in question. The fundamental problem in the analysis of porous systems is that it is only the spatial change of the scattering length density which is important, not its sign. In other words, one may describe the same morphology as scattering caused by the matrix (with holes), or by the holes themselves (identified by the surrounding matrix) [20] – in optics this is called Babinet's principle. The usual approach is to describe the minority component geometrically, although symmetric bicontinuous microemulsions are a striking counterexample. In the present study with a focus on pores, the scattering in absolute units will be analyzed in terms of the shape of the pores, based on quantitative information from BET on their number, width, and volume. SAXS and sorption isotherms have been shown to give comparable results with respect to pore size in ordered mesoporous silica. [40]

A major drawback of monoliths is their powder-like appearance which makes it difficult to cast them into a desired shape. In dielectric studies of the conducting properties, e.g., a flat and homogeneous sample of well-defined thickness is necessary for quantitative evaluations. Similar issues arise with small-angle scattering techniques, which are usually based on thin specimen in order to be able to reason in absolute units. The most significant progress of the present study arises therefore from the development of a new type of ionosilica, which we call 'ionosilica films' because of their flat and well-defined geometry. It has been obtained by a hydrolytic synthesis strategy.

In this article, we compare this new hydrolytic way of synthesizing porous ionosilica to the well-established non-hydrolytic one. [30-32] The starting point is the meso- and macroporosity of both systems as measured by a combination of TGA and nitrogen sorption. This is followed by a SAXS study for which we propose a geometrical model based on the BJH pore polydispersity in width capable of predicting the scattering in absolute units. The combination of all these results will be represented in a unified way, highlighting the higher porosity of the hydrolytic films with respect to monoliths obtained via a non-hydrolytic sol-gel procedure. Our analysis shows that the new synthesis protocol of films allows the formation of mesoporous systems of virtually identical pore geometry, but with a higher incorporation of ionic liquid, and thus higher mesopore volumes.

**Materials and methods**

**Chemicals.** Formic acid and hydrochloric acid were purchased from VWR and used as received. The imidazolium-based ionic liquid, 1-butyl-3-methylimidazolium bis(trifluoromethylsulfonyl)imide (BMIM-TFSI) was purchased from IoLiTec. The ionosilica precursor tris(3-(trimethoxysilyl)propyl)amine (TTA) was synthesized following previously described protocols. [6]

**Syntheses.** The synthesis of ionosilica materials was performed in a one-pot synthesis involving two main components, namely TTA and BMIM-TFSI. Ionosilica matrices were formed in two different ways. The first is based on a non-hydrolytic sol–gel process that produces 'monoliths', while the second is a newly developed hydrolytic sol-gel process resulting in 'films'. The difference between the two terminologies – film and monolith – lies in the fact that the sample geometry is well controlled for the films, but not for the monoliths, which are powder-like materials made of agglomerated grains. The non-hydrolytic sol-gel process is outlined elsewhere.[32] Briefly, the TTA precursor (3 ml) was mixed with various amounts of BMIM-TFSI (between 1 and 12 ml), and the polycondensation was initiated by adding 2 ml of formic acid. Gelation results from the transesterification–polycondensation reactions of the trimethoxysilyl groups of the TTA precursor, the protonation of which provides the ionic building blocks of the IS matrix. It typically took three days to complete and was conducted in closed glass test tubes. On the other hand, the hydrolytic film synthesis was performed by mixing TTA (3 ml) and BMIM TFSI (from 1 to 12 ml) for a few minutes, and then adding 3 mL of ethanol. The polycondensation reaction was initiated by adding 0.65 ml of a 0.6 M hydrochloric acid solution. The obtained mixture was poured into a closed Teflon mold drilled with two small holes on the upper side and stored at room temperature for 48 hours, resulting in a homogeneous film of approximately 1 mm thickness.

In both hydrolytic and non-hydrolytic processes, the porosity of the ionosilica scaffold is produced by *in situ* templating with the ionic liquid. Since all other quantities are kept fixed, the volume of IL introduced ($v_{IL}$) is the main experimental parameter that identifies the samples. For a given quantity of precursor (3 ml), four values of $v_{IL}$ were chosen: 1, 3, 6 and 12 ml. TGA will be used to convert them to the final volume fractions $\Phi_{IL}$ of ionic liquid in the samples. A total of 25 samples (see Tables S1 and S2 in ESI) were prepared to test for reproducibility. Once the materials are formed, the ionic liquid can be easily extracted by washing as shown in Scheme 1a, and $\Phi_{IL}$ is thus identical to the total pore volume fraction. In practice, each sample was divided into two parts: one part was kept as such (without rinsing), while the other part was 'washed' using a Soxhlet apparatus with ethanol for 48 h, to remove IL from the pores. The materials were finally dried under reduced pressure (10 mb) at 80°C overnight (Scheme 1c).

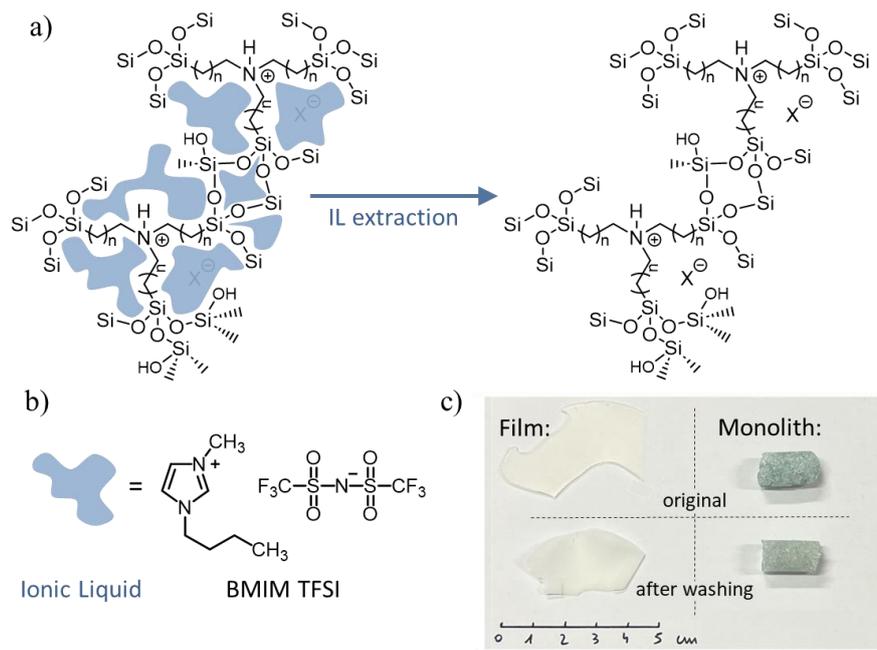

**Scheme 1:** (a) Atomic structure of the ionosilica scaffold before (left) and after (right) removing the confined IL by washing with ethanol. X⁻ is the counter ion corresponding to Cl⁻ and COO⁻ for the hydrolytic and non-hydrolytic sol-gel process, respectively. (b) Atomic structure of the ionic liquid. (c) Photographs of the resulting film and monolith ($v_{IL}$ = 6 ml).

**Thermogravimetric Analysis (TGA).** The IL fractions were obtained by TGA (Mettler Toledo, 10 K/min under air) from the weight loss of the composite samples between 120 and 1000°C leading to the inorganic residue ($SiO_2$). Knowing the molar ratio between $SiO_2$ and ionosilica ($n_{SiO2}/n_{IS}$ = 3, see Scheme 1), one can deduce the IS mass and thus the IL weight fraction. The volume fractions were determined by mass conservation using the densities of IS ($d_{IS}$ = 1.58 g/cm³ by pycnometry [31]) and IL ($d_{IL}$ = 1.43 g/cm³ by using a densimeter).

The TGA curves for pure IL, IS/IL and washed IS/IL are given in ESI (Figure S1). They show a sharp degradation of the pure IL around 450°C as also observed in IS/IL composites, whereas it is significantly broadened in IS/IL after washing. The optimal extraction time of two days for IL was determined with a series of monoliths and films ($v_{IL}$ = 6 ml). The samples were washed with anhydrous ethanol for increasing times ranging from 10 minutes in a beaker up to 5 days in Soxhlet. The lower weight loss measured by TGA and the increasing ionosilica contribution at high-q in SAXS show that all incorporated IL has been extracted after two days (see Figures S2 and S3 in ESI). Moreover, there is no detectable glass transition in the calorimetric measurements of the IS/IL washed samples.

In the following, we have thus excluded the presence of any significant residual IL amount in the washed samples. From the weight-loss of pure ionosilica, one can estimate its molecular weight: $MW_{IS}$ = $m_{IS}/n_{IS}$ = 3 $MW_{SiO2}$ $m_{IS}/m_{SiO2}$, where $m_{IS}$ correspond to the initial mass taken at 120°C to account for unavoidable water uptake and $m_{SiO2}$ is the residue mass. $MW_{IS}$ is found to be systematically above the molecular weight of the fully condensed molecule (ca. 110%) indicating that the condensation of methoxy groups of the TTA precursor is not complete. On average over the different samples, about 2.5 methoxy groups over 9 did not condensate. It should be noted that the IS molecular weight thus obtained (and not the theoretical value) was considered to determine the IL content of the composite sample before washing.

**Nitrogen sorption.** $N_2$ adsorption-desorption isotherms at 77.3 K were obtained using a Micromeritics TriStar volumetric apparatus after outgassing the samples at 120 °C under reduced pressure (10 mb) overnight. The isotherms give the adsorbed volume of $N_2$ gas per unit sample mass as a function of the reduced pressure $p/p_0$. The fit in the BET domain (0.05 < $p/p_0$ < 0.35) provides the BET parameters C associated with the shape of the adsorption isotherm, and $v_m$. The latter corresponds to the monolayer coverage in $cm^3/g$. The point on the adsorption curve associated with the monolayer coverage is called B. This point provides an estimate of the specific surface per unit sample mass, $S_{BET}$. $S_{BET}$ is directly proportional to $v_m$

$$S_{BET}(m^2/g) = \frac{s\, N_A}{v_{mol}} v_m = 4.36\, v_m\, (cm^3/g) \tag{1}$$

where $v_{mol}$ = 22.4 L/mol is the molar volume of gaseous nitrogen, s = 0.162 $nm^2$ its surface per molecule, and $N_A$ the Avogadro constant. The total BET pore volume $v_{pore}$ per unit sample mass has been estimated from the high-pressure plateau of the adsorbed gaseous amount $v_{max}$,

$$v_{pore} = \frac{d_{gaz}}{d_{liq}} v_{max} \tag{2}$$

where $d_{gaz}$ and $d_{liq}$ refer to the density of gaseous (0.00125 $g/cm^3$) and liquid nitrogen (0.808 $g/cm^3$), respectively.

The combination of BET and TGA results has been used to determine volume fractions of ionosilica ($\Phi_{IS}$), ionic liquid ($\Phi_{IL}$), and mesopores ($\Phi_{meso}$). Note that data points with error bars correspond to averages of different runs including a few data points previously published by us [32]. By conservation of volume, the following sums of volume fractions are known:

$$\Phi_{IS} + \Phi_{IL} = 1 \tag{3a}$$

$$\Phi_{IS} + \Phi_{meso} + \Phi_{macro} = 1 \tag{3b}$$

Eq.(3a) corresponds to the result of the TGA-measurement, which allows differentiating ionic liquid from ionosilica, and eq.(3b) to the one of adsorption measurements providing $v_{pore}$. The latter can be converted into the volume fraction of mesopores using $\Phi_{IS}$ value from TGA:

$$\Phi_{meso} = v_{pore}\, d_{IS}\, \Phi_{IS} \tag{4}$$

$\Phi_{macro}$ in eq.(3b) denotes the volume fraction of macropores which are out of range for BET, i.e., too big. All pores considered here are filled with ionic liquid after synthesis and become visible in SAXS only after washing. This implies that all pores are open.

The size of the pore can be estimated using a simple calculation for pores taken as cylinders of infinite length and monodisperse diameter. Then, the ratio of the pore volume $v_{pore}$ and the specific surface $S_{BET}$ gives an estimate of the pore diameter: $D_{pore}$ = 4 $v_{pore}/S_{BET}$. The origin of this equation is easy to understand, as $v_{pore}$ is proportional to $v_{max}$, and $S_{BET}$ to $v_m$: The higher the ratio, the more volume is available, and thus the bigger the average diameter. The ratio in height between the high-pressure plateau and the point B related to the monolayer coverage thus provides a quick estimation of the average pore diameter in nanometers: $D_{pore}$ = 1.44 $v_{max}/v_m$.

**Transmission Electron Microscopy (TEM).** High resolution TEM pictures were recorded with a JEOL 2200FS apparatus at 200 kV from MEA platform, Université de Montpellier (France). Monoliths embedded in a resin and films were cut into slices of nominal thickness e = 70 nm and were placed on a metal grid.

**SAXS.** Small-angle X-ray measurements were performed on two different devices. An in-house setup of the Laboratoire Charles Coulomb (L2C), "Réseau X et gamma", Université de Montpellier (France) was employed using a high-brightness low-power X-ray tube with a wavelength λ = 1.54 Å, coupled with aspheric multilayer optics (GeniX3D from Xenocs). The differential scattering cross sections (called intensities for simplicity) were measured with a 2D "Pilatus" pixel detector at two sample-to-detector distances d = 1830 and 180 mm, leading to a q range of $6.3 \times 10^{-3}$ to 1.9 Å$^{-1}$. Other SAXS experiments were conducted on beamline SWING at synchrotron SOLEIL (Saint Aubin, France) using the two following configurations: 12 keV with d = 6230 mm and 15 keV with d = 520 mm, giving a q-range from $10^{-3}$ to 2.4 Å$^{-1}$). Standard data reduction tools provided by Soleil were used (Foxtrot 3.1). SAXS intensity values in absolute units were obtained by comparing to a secondary standard (water at Soleil, high-density polyethylene at L2C).

The absolute units of the scattering cross sections provide valuable information on the number and volume of pores, but the thickness of the sample needs to be known. The determination of the exact thickness at the position of impact of the X-ray beam, however, remains tricky, and induces some fluctuations in height between the different samples. We therefore have normalized all curves of the film samples after washing to their ionosilica content, $I(q)/\Phi_{IS}$, and superimposed them at high q, in the WAXS range between 0.4 and 2.4 Å$^{-1}$, where the large-angle peaks reflect local correlations between ionosilica atoms. The reference high-q level for the superimposition was estimated from the average of five film samples of the same composition (6 ml of IL). The monoliths after washing were in the form of powder measured in glass capillary. They were all rescaled to the high-q film intensity in $I(q)/\Phi_{IS}$. Both films and monoliths before washing were normalized to their IL content, $I(q)/\Phi_{IL}$, and rescaled to the high-q intensity of the pure IL measured in glass capillary.

**SAXS analysis.** The spatial modulation of the scattering length density gives rise to the scattering. For inhomogeneous solids, Debye has proposed a description based on a single correlation length ξ predicting a strong decay in q following the Guinier regime. [41,42] This decay is compatible with our data but the intensity prefactor is far too weak. Alternatively, the scattering can be described by models of either matrix ionosilica (e.g., thought of a construction of primary spheres), or by a model of pores in that matrix. To account for both intensity value and shape, we have designed a model that interprets the scattering as coming from the morphology of cylindrical pores. It will be shown that this picture is compatible with an increase of pore size with the quantity of templating IL, whereas the complementary picture would lead to an increase in size of the primary matrix constituents with templating liquid (see ESI), which would be a highly unexpected scenario. In the IL/IS composites, the ionosilica is thus considered as the matrix and the scattering contrast of the IL-filled pores is given by $\Delta\rho = \rho_{IS} - \rho_{IL} = 1.7 \times 10^{10}$ cm$^{-2}$, with $\rho_{IS} = 14.3 \times 10^{10}$ cm$^{-2}$ and $\rho_{IL} = 12.5 \times 10^{10}$ cm$^{-2}$ determined from the chemical formula and the respective density. Such a contrast value is extremely low, and it does not allow the quantitative characterization of the pore scattering by SAXS. This is not the case in the emptied samples after washing. The latter contain ionosilica and pores of different sizes – namely meso- and macropores – the proportions of which are described by eqs.(3). There, the contrast of the empty pores is significant, given by $\Delta\rho = \rho_{IS} - \rho_{air} = 14.3 \times 10^{10}$ cm$^{-2}$.

The total scattered intensity then reads:

$$I(q) = \Phi_{macro} I_{macro}(q) + \Phi_{meso} I_{meso}(q) \qquad (5)$$

We will see in the results section that the big macropores are outside the q-window of the SAXS experiment, thus $I_{macro}(q) = 0$ and all small-angle scattering stems from the mesopores. The quantity of the latter is known from TGA and BET, and the shape of the scattering curves in the low-q range will be analyzed in detail below. It appears to be compatible with cylindrical pores, the form factor of which for a length $L_{cyl}$ and a circular cross section of radius R is given by [43,44]

$$P(q) = \int_0^{\pi/2} \left[ \frac{2J_1(qR\sin\alpha)}{qR\sin\alpha} \frac{\sin(qL_{cyl}\cos\alpha/2)}{qL_{cyl}\cos\alpha/2} \right]^2 \sin\alpha \, d\alpha \qquad (6)$$

where the integral over α performs a numerical orientational average for isotropic distributions of directions of the cylinders. $J_1$ is the first order Bessel function. For monodisperse cylinders, the SAXS intensity $I_{meso}(q)$ as defined in eq. (5) is simply obtained by multiplying the form factor by the scattering contrast squared $\Delta\rho^2$ and the volume of the cylinder $V_{cyl} = \pi R^2 L_{cyl}$.

In presence of polydispersity in cylinder radius, the radius obeys a distribution function $N_{cyl}(R) \, dR$, which gives the number of cylinders per unit volume of radius between R and R + dR. Multiplying this number by the cylinder volume $V_{cyl}$ gives the contribution of pores of this width to the pore volume fraction. The novelty of our approach is that this quantity is directly determined from the stepwise adsorption of the nitrogen for each pore width fulfilling the Kelvin equation, if the surface adsorbed layer is taken into account as done by the BJH theory. [37] We thus introduce the size distribution and total quantity of pores obtained by BJH from the adsorption isotherms into the quantitative description of the SAXS polydispersity and thus intensity in absolute units.

The scattered intensity for independent cylinders with radial polydispersity then reads:

$$I_{cyl}(q) = \Delta\rho^2 \int N_{cyl}(R) \, V_{cyl}^2 \, P_{cyl}(L_{cyl}, R) \, dR \qquad (7)$$

Integrating $N_{cyl}(R)$ over all pore widths gives the total cylindrical mesopore volume fraction:

$$\int N_{cyl}(R) \, \pi R^2 L_{cyl} \, dR = \Phi_{cyl} = \beta \, \Phi_{meso} \qquad (8)$$

The parameter $\beta \in [0, 1]$ allows to adjust the contribution of the cylindrical pores to the mesopore volume fraction, the rest (1-β) being discussed below. The length of the pores $L_{cyl}$ entering eq.(7) has been fixed by fitting of the intensity decay for all pores of a given sample. The pore length needs to be further discussed, as on first sight the BJH and the SAXS models appear to be incompatible. The BJH pores are thought to be infinite in length, which however expresses only the fact that the radius is the key parameter: they are filled whenever their radius meets the Kelvin radius, and if there are no other complications like bottlenecks to be described. The total cylinder length in a sample is obviously not infinite, and in fact one can divide the finite pore volume measured by adsorption dV (of pores of a given radius between R and R+dR) by their surface and one obtains the total length of such pores present in the sample. In other words, BET and BJH implicitly describe a single pore of finite length for each radius. By dividing this total length by the SAXS cylinder length $L_{cyl}$, or equivalently, by dividing the volume by the cylinder volume, one obtains the number of cylindrical segments of length $L_{cyl}$ (and radius R) seen by the SAXS intensity, and thus $N_{cyl}(R)dR$, the number per unit sample volume:

$$N_{cyl}(R)dR = d \, \beta \, dV / V_{cyl} \qquad (9)$$

where d = $d_{IS}\Phi_{IS}$ is the density (g/cm$^3$) of the porous material, and dV is the output of the BJH analysis, i.e., the pore volume per unit sample mass of cross section radius between R and dR. In standard BJH analysis, the function dV/dR is usually plotted, and one sees from eq.(9) that it is equal to the number density $N_{cyl}$ multiplied by the volume of the corresponding pore, in other words the contribution to the volume fraction. This volume fraction interpretation will be used in this article to represent the polydispersity. Here β takes explicitly into account that not all BJH volume is in cylindrical pores. The two representations of spatial structure are compared in Scheme 2.

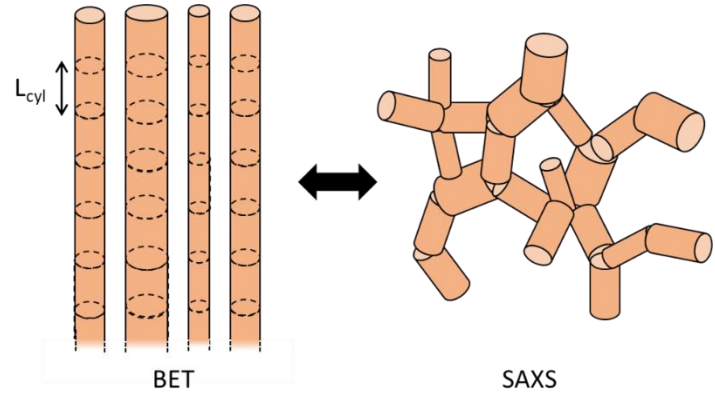

**Scheme 2:** (a) BJH image of infinite cylindrical pores of different diameters). (b) SAXS image of small pieces of cylinder of single length $L_{cyl}$ connected randomly. Note that the number of cylindrical sections is identical in a) and b), as in the model.

When fitting the cylinder contribution to the SAXS intensity, the high-q data (0.1 – 0.2 Å$^{-1}$) is not correctly reproduced. This is presumably due to the fact that the cylinders do not have ideally smooth walls, nor perfectly constant radii. The situation can be improved somewhat by adding some smaller substructure to the cylinder scattering, representing these imperfections. We have tentatively added the scattering of small polydisperse spheres, of radii obeying a Gaussian distribution function, and comparable to the cylinder radius. This second contribution represents a part 1-β of the mesopore volume fraction, and it scattered intensity reads:

$$I_{sphere}(q) = \Delta\rho^2 \int N_{sphere}(R)\, V^2_{sphere}(R)\, P_{sphere}(R)\, dR \qquad (10)$$

$P_{sphere}(q)$ the polydisperse form factor of spheres normed to 1 at small angles, and $V_{sphere}$ is the volume of the spherical pores of radius R. [38] Integrating $N_{sphere}(R)$ over all spheres gives the total sphere contribution to the mesopore volume fraction:

$$\int N_{sphere}(R)\, \frac{4\pi}{3} R^3\, dR = \Phi_{sphere} = (1-\beta)\,\Phi_{meso} \qquad (11)$$

By construction, the total mesopore volume fraction remains conserved and equal to the output of the adsorption isotherms:

$$\Phi_{meso} = \Phi_{cyl} + \Phi_{sphere} \qquad (12)$$

The total scattering function thus reads:

$$I(q) = \Phi_{meso}\, I_{meso}(q, \beta) = I_{sphere}(q, \beta) + I_{cyl}(q, \beta) \qquad (13)$$

where we have explicitly recalled the dependence on the parameter β – the contribution to scattering fraction of cylinders and spheres, respectively, among all mesopores. It will be dropped later in the text.

**Results and discussion**

Film and monolith samples have been synthesized as described in the methods section. They have been characterized by TGA to quantify the amounts of ionic liquid and ionosilica. The ionosilica residue at high temperature (i.e., silica) has been measured as a function of sample composition, for both systems, hydrolytic and non-hydrolytic samples, before washing. These measurements serve to evaluate the amount of ionic liquid effectively introduced in the systems, with respect to the nominal amount, and can thus be interpreted as the efficiency of IL-incorporation into the ionosilica. The TGA curves have been normalized to the mass at 120°C to account for water uptake. The degradation of the ionic liquid in ionosilica is observed around 450°C, as for the pure IL. The two sets of composite samples have also been measured after washing, to estimate the average degree of condensation of methoxy groups as explained in the methods section. Moreover, the TGA curves show that repeating experiments with different samples of same nominal composition offer very good sample reproducibility.

In Figure 1, the incorporated IL volume fraction deduced from TGA measurements is plotted. This graph shows that the IL quantity in the composites increases with the nominally introduced IL quantity for all samples. By comparing to the maximum IL volume fraction based on the nominally introduced quantities (see the methods section), we can determine the efficiency of incorporation. The corresponding fractions in terms of successfully incorporated IL are given in the ESI (Figure S4). Figure 1 shows that the ionic liquid is fully incorporated into the porous ionosilica for films across the entire parameter range tested in our experiments. Indeed, the expected and the measured IL-volume fractions agree within error bars. For the non-hydrolytic monolithic samples, the efficiency is lower. The latter increases monotonously from ca. 30 to 90%, and for 6 ml samples, it is about 75%. This means that a fraction of the ionic liquid is not incorporated in the ionosilica matrix during the monolith synthesis leading to more syneresis than in the films. The IL incorporation is thus dependent on the synthesis routes, being favored (up to 100%) in the case of a hydrolytic sol-gel process. When discussing the mesostructure of the samples as seen by SAXS below, the volume fraction occupied by the templating ionic liquid – which becomes the volume fraction of empty pores after washing – will be a natural parameter to describe the sample properties. This corresponds to a change in variables away from the experimental parameter $v_{IL}$ on the x-axis of Figure 1 to the resulting $\Phi_{IL}$ on the y-axis, thus integrating the characterization by TGA in the description. The advantage of $\Phi_{IL}$ is that it gives directly the relevant final sample composition, which is necessary in particular for the SAXS analysis.

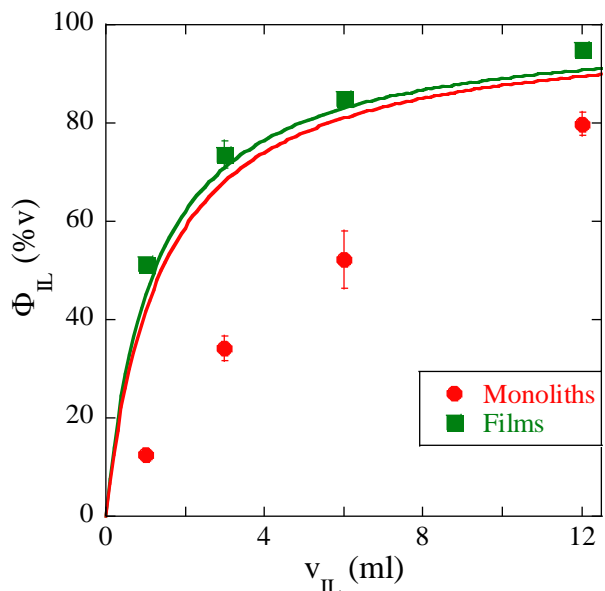

**Figure 1:** Volume fraction of effectively incorporated ionic liquid into the ionosilica scaffold measured by TGA as a function of the nominal amount of ionic liquid introduced, for the two synthesis routes (monoliths and films). This quantity is expressed as the final volume fraction of ionic liquid in the sample. The expected IL volume fraction assuming full incorporation is calculated from the nominal amounts in both cases (solid lines).

The TGA results provide quantitative information on the total volume fraction of pores present in both the monoliths and the films as shown in Figure 1. In order to probe these pores by BET, they have been emptied by washing following the procedure described in the methods section. Anticipating the SAXS results to come later in this article, empty pores contribute to the SAXS at intermediate scattering vectors, while filled ones do not, due to the lack of contrast between ionosilica and ionic liquid. The existence of a strong SAXS signature thus establishes that the pores are interconnected and open. If some closed pores exist, they contribute neither to the SAXS nor to sorption isotherms, and they can thus be safely disregarded in the present article.

Nitrogen adsorption-desorption isotherms have been measured to determine the specific pore surface and the amount and size of the pores. A series of isotherms for both monoliths and films is shown in Figure 2 in terms of adsorbed volume per sample mass vs the relative pressure. By comparing the two families of isotherms for monoliths and films, it appears that the adsorbed quantities are clearly higher for the films. It also increases when increasing the nominal IL quantity from 1 to 12 ml. It can be concluded that the amount of IL and the synthesis route allow tuning the specific surface, as well as the amount and size of the mesopores. Monoliths at the lowest IL-fractions do not show any porosity and thus no adsorption: the signal for 1 ml sample is zero, and for 3 ml sample it is very low as shown in Figure 2. Only for larger ionic liquid contents, monoliths possess porosity. This is a striking difference to the hydrolytic systems, which present porosity from the lowest IL content on. As a result, the adsorbed $N_2$-quantities are always higher for the latter group.

Besides the non-porous monoliths, all other samples show the presence of micropores of typical diameter below 2 nm at very low reduced pressure, and then a strong contribution from mesopores in the typical BET range from $p/p_0$ = 0.05 to 0.35. The general shape of the isotherm can be classified as type IV, [36] with the presence of an inflection point B around $p/p_0$ = 0.1 indicative of the saturation of the surface by the monolayer. At intermediate pressures, pore filling is observed, with a hysteresis loop due to more difficult pore emptying during desorption possibly caused by slow nitrogen evaporation through bottlenecks in or between the pores. At the highest pressures, a plateau showing complete pore filling is visible, in some cases accompanied by an upturn indicative of the existence of macropores, which however are out of range for this technique. Indeed, as with the SAXS analysis of the Porod domains provided below, macropores are at least 1000 times larger than the mesopores observed here.

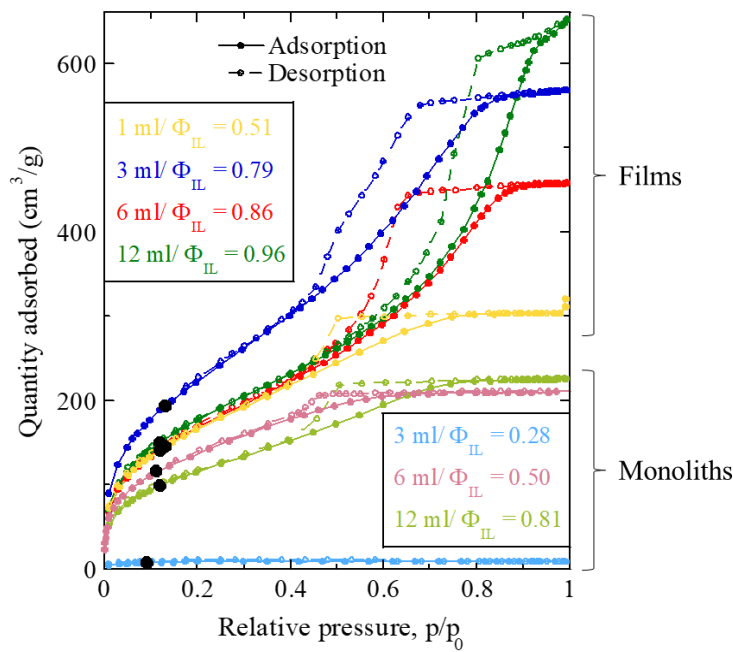

**Figure 2**: Nitrogen adsorption-desorption isotherms at 77.3 K for series of monoliths and films as indicated by the labels, for increasing amounts of introduced IL as given in the legend. The black point on each curve represents point B.

The analysis of the specific surface can be performed by fitting the BET equation [45] to the low/intermediate pressure range ($p/p_0 < 0.35$) – which in practice is done by plotting $[V(p_0/p-1)]^{-1}$ as a function of relative pressure. The result is a straight line in this range as shown in the ESI (Figure S5), and the BET parameters $v_m$ and C can be deduced. Together with other parameters, they are reported in Table 1. The C-parameter stays roughly in the 50 to 100 range, indicating similar shapes of the adsorption isotherm as found by visual inspection. The adsorbed monolayer volume $v_m$ is reached at $p/p_0 = [\sqrt{c} + 1]^{-1}$, which is typically 0.1 for our data as also shown in Table 1 and indicated by the position of the point B in Figure 2. As described in the methods section, the specific surface can be deduced from $v_m$ and molecular parameters. For films, the monolayer coverage $v_m$ is found to be in the 150 – 200 cm$^3$/g range, with a maximum reached for the 3 ml sample. This translates into specific surfaces $S_{BET}$ of around 650 m$^2$/g, with a maximum at 850 m$^2$/g. As one can expect from the monolayer coverages, there is considerably less mesoporosity in monoliths than in films. The monolayer coverage in monoliths is unmeasurably low for 1 ml sample, and then increase up to about 100 cm$^3$/g. Correspondingly low $S_{BET}$ values below 450 m$^2$/g are found, a value reached only for the highest IL content of 12 ml.

**Table 1:** Analysis of adsorption isotherms of films and monoliths. Nominal IL content, incorporated IL volume fraction, maximum adsorbed gas quantity $v_{max}$, BET fit parameters C and adsorbed gas volume $v_m$ at monomer coverage, specific surface, total pore volume, average pore diameter, relative pressure at monolayer coverage.

|  | $v_{IL}$ (ml) | $\Phi_{IL}$ (%v) | $v_{max}$ (cm³/g) | C (fit) | $v_m$ (fit) (cm³/g) | $S_{BET}$ (m²/g) | $v_{pore}$ (cm³/g) | $D_{pore}$ (nm) | $p/p_0$\|mono |
|---|---|---|---|---|---|---|---|---|---|
| Films | 1 | 51.4 | 304 | 56 | 141 | 615 | 0.47 | 3.1 | 0.12 |
|  | 3 | 78.5 | 566 | 46 | 194 | 843 | 0.88 | 4.2 | 0.13 |
|  | 6 | 85.7 | 457 | 48 | 145 | 630 | 0.71 | 4.5 | 0.13 |
|  | 12 | 95.8 | 642 | 52 | 151 | 658 | 1.00 | 6.1 | 0.12 |
| Monoliths | 1 | 15.8 |  |  |  |  |  |  |  |
|  | 3 | 27.6 | 9 | 96 | 8 | 33 | 0.01 | 1.8 | 0.09 |
|  | 6 | 50.2 | 210 | 62 | 116 | 504 | 0.33 | 2.6 | 0.11 |
|  | 12 | 79.3 | 225 | 58 | 99 | 429 | 0.35 | 3.3 | 0.12 |

For infinite pores which are cylindrical and monodisperse in diameter, the average pore diameter can be deduced (see methods section) from the ratio of $v_{max}$ to $v_m$. As given in Table 1, these diameters evolve monotonously from 3 to 6 nm with increasing IL content for films, and from 2 to 3 nm for monoliths, which have thus thinner pores.

More elaborate theories like density functional theory or BJH [35] can also be applied in order to estimate pore size distribution functions. BJH has been used to estimate the pore sizes corresponding to each adsorption step, and the width of the size distribution function has been successfully used as input in the SAXS analysis below. Some results are given in the ESI (Figure S6) for the sake of completeness.

The total volume fraction of mesopores $\Phi_{meso}$, as seen by adsorption, can be determined from the plateau value of the isotherms, $v_{max}$, which gives the specific pore volume $v_{pore}$ following eq.(2). The average values over several measurements of $\Phi_{meso}$ have been plotted together with the TGA data for $\Phi_{IS}$ in Figure 3, for both films (3a) and monoliths (3b). The volume fraction of mesopores is found to be lower for monoliths than for films at low IL fractions, indicating that films have superior pore forming properties. Indeed, it has been shown in Figure 1 that hydrolytic films incorporate all IL, and mesopore volume fractions of up to 40% are obtained at low IL content. Adding more ionic liquid does not help the formation of mesopores; we will see below that additional IL is essentially converted into macropores. In terms of mesopore formation, the best parameters are thus those generating hydrolytic films with little IL. In both monoliths and films, it is observed that the ionosilica volume fraction $\Phi_{IS}$ decreases with the sample composition – increasing IL-content –, starting from the nominal value of 100% in absence of ionic liquid. The decay can be described by a simple exponential in the case of monoliths, and this function is superimposed to the data as a guide to the eye in Figure 3b. For films, however, the decay is considerably steeper, and a stretched exponential had to be used to describe the data points in Figure 3a. As indicated in the discussion of Figure 1, it is helpful to switch to a description in terms of the final pore volume fraction given by $\Phi_{IL}$. The corresponding graphs are presented in Figure S7. Then $\Phi_{IS}$ has the trivial dependence on $\Phi_{IL}$ given by eq. (3a), and the graph highlights the contributions of meso- and macropores, respectively, which we investigate now.

Using the conservation of volume expressed by eq.(3b), there must be a remaining volume fraction of pores created by the ionic liquid which is out of range of the BET measurements. This volume fraction $\Phi_{macro}$ corresponds to macropores, and it has been calculated by subtraction. It is also plotted in Figure 3. $\Phi_{macro}$ is found to increase with the introduced IL-quantity, for both synthesis routes. It thus appears that most of the ionic liquid which is added in the synthesis generates macropores at high IL content. On the other hand, the amount of mesopores remains below some 20% in volume for most samples – only films at low IL content (< 3 ml) approach a high mesopore volume of 40%. These films thus have superior mesoporosity as compared to monoliths. Another way of expressing the same tendency is to say that in the case of monoliths, there is generally more ionosilica than mesopores in the samples, whereas the fraction is comparable for most hydrolytic film samples.

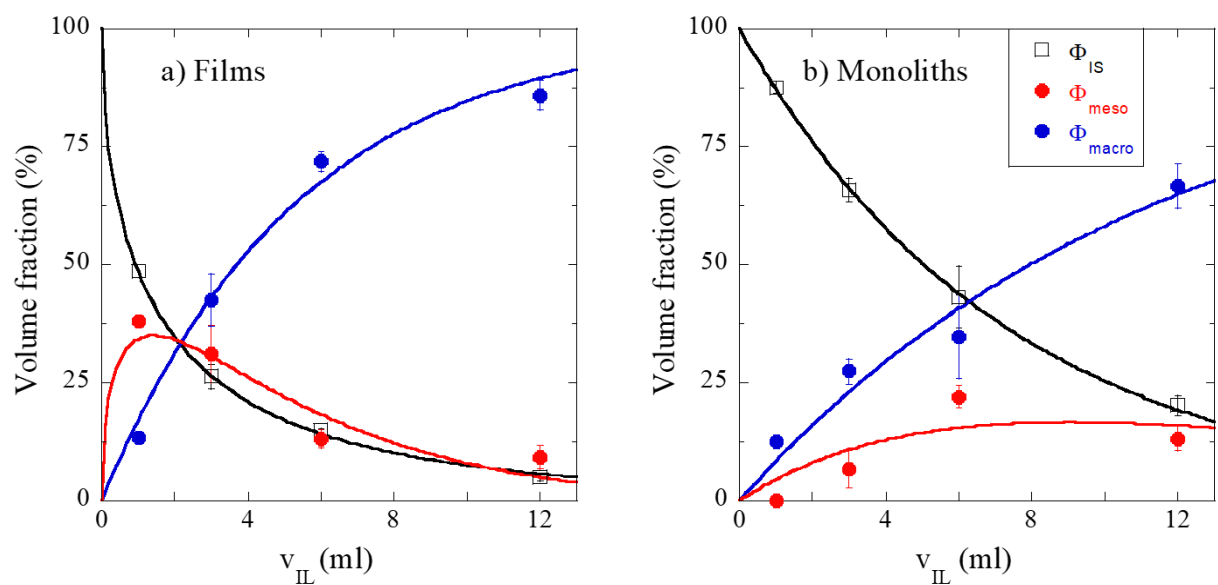

**Figure 3:** Volume fractions of ionosilica and mesopores in (a) films and (b) monoliths, determined by TGA and BET, as a function of the introduced IL volume. The remaining volume fraction $\Phi_{macro}$ corresponds to pores which are too big to be captured by BET. Lines are guides to the eye respecting the conservation of volume given by eq. (3b).

The higher porosity of the films compared to monoliths may result from different reaction mechanisms during the sol-gel transformation. Indeed, a large quantity of synerese liquid is formed during the formation of the monoliths, due to shrinkage of the monoliths and expulsion of liquid from the materials. This liquid is composed of solvent (formic acid), reaction by-products such as ethyl formate, but also of a certain volume fraction of ionic liquid that is therefore lost during the sol-gel transformation. On the contrary, the whole amount of IL is kept within the material during the formation of films. This behaviour explains (*i*) the higher IL content of the films compared to the monoliths on the one side and (*ii*) the higher porosity of the films after IL elimination via washing.

Macroporosity is also evidenced by TEM. An example is shown in Figure 4 for the 1 ml composites. Surprisingly, one can see that a well-defined structure has been formed in both film and monolith (it is also present at higher IL content, see Figure S8). Macroporosity has the appearance of a lamellar structure with a periodicity of about 200 and 300 nm for the film and monolith, respectively. Such a structure may be the result of a supramolecular organization of the TTA precursor and the part of the ionic liquid which is not in the mesopores. Since this phenomenon has not been observed in other non-ionic systems so far, we believe that it may be induced by specific interactions between the precursors and the IL during the sol-gel transformation. These interactions may result in phase segregation and to the formation of lamellar morphologies as shown in Figure 4. Mesoporosity is well visible in the film at higher magnification with a disordered wormlike structure and characteristic pore sizes in the range from 5 to 10 nm. It is however difficult to conclude on any pore shape due to the superposition of different layers of material (nominal thickness of ca. 70 nm), which results in a low contrast in transmission. In agreement with the BET results, the mesoporosity of the 1 ml monolith is not visible in the TEM pictures.

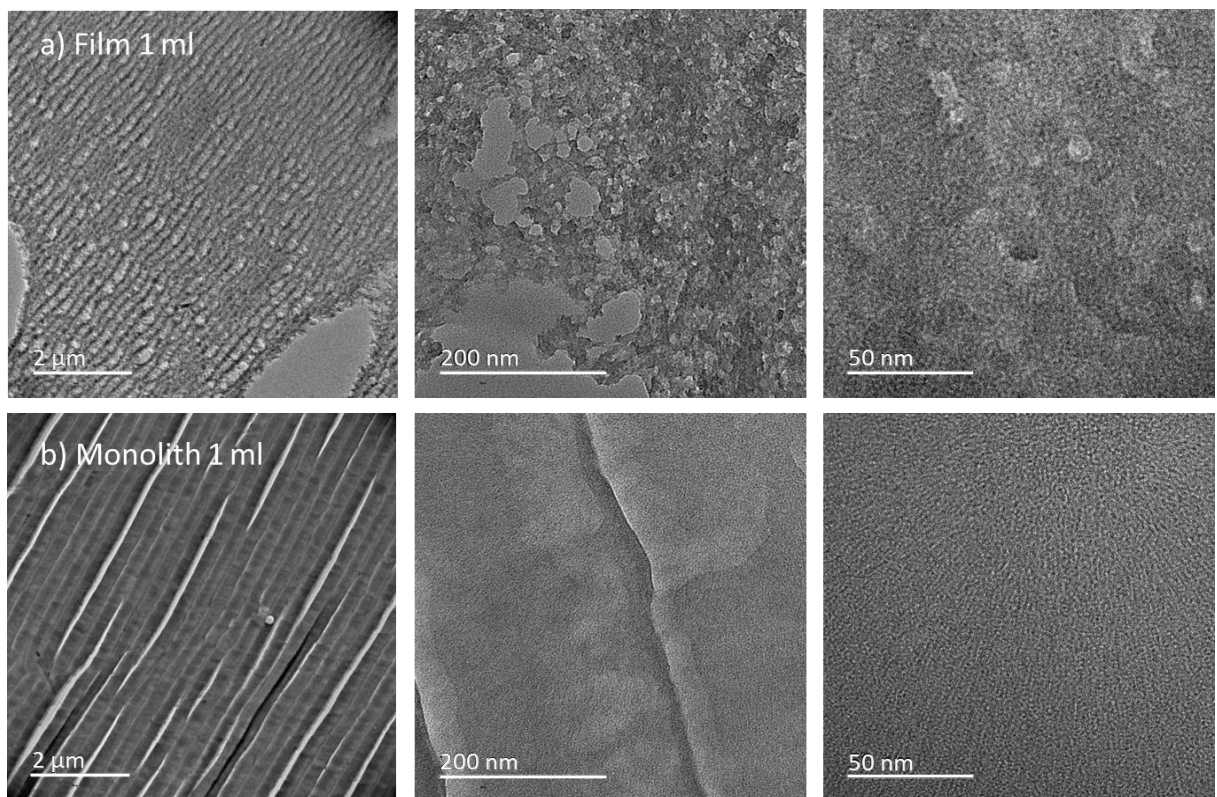

**Figure 4:** High-resolution TEM pictures of ionosilica composites after Soxhlet washing for three magnifications. (a) Film and (b) monolith. Both were prepared with 1 ml of ionic liquid, $\Phi_{IL}$ = 51.4%v for the film and 13.0%v for the monolith.

The size of the mesopores has been characterized by SAXS. We exemplarily start the discussion with the scattering of the different constituents. In Figure 5a, the intensity of a 6 ml-film still filled with ionic liquid is compared to the one of the same sample after washing. As discussed in the methods section, the scattering contrast between IL and the ionosilica matrix is very low, and pores are essentially invisible before washing. The only strong scattering feature is a low-q upturn, which is presumably due to the presence of crazes or other large-scale heterogeneities. At large q, the atomic correlations of the ionic liquid are found, as evidenced by the superposition with the pure IL. At low angles, the IL scattering is seen to stay weak due to the large-scale homogeneity of the liquid. Once the pores are emptied, a prominent SAXS signal arises: In Figure 5a, the intensity curve of the emptied film evidences the presence of structure on a nanometric scale, with a well-defined plateau followed by a Guinier decay at around 0.04 Å$^{-1}$. Pore emptying was thus successful, which also indicates that pores are connected as shown in Scheme 2. One may note that the weak signal in the intermediate q-range of IL-filled ionosilica positioned right to the arrow labelled "before washing" in Figure 5a is compatible with the typical size of the pores visible in the emptied sample, at however a much weaker contrast.

We have several examples of the effect of washing, for different amounts of ionic liquid. In Figure 5b, the equivalent result for washing of monoliths with a higher amount of IL (12 ml) is presented, showing again that washing allows highlighting the mesostructure. Also, we have superimposed the scattering of the pure ionic liquid for comparison to the filled sample (i.e., before washing). Once the samples are emptied, the scattering peaks in the WAXS region are only due to the ionosilica: the heights are by construction in agreement with the IL and IS concentrations given in Figure 1: There is a bit less ionosilica in the 6 ml film (14%) (and more IL, 86%) with respect to the 12 ml monolith (19% IS, 81% IL).

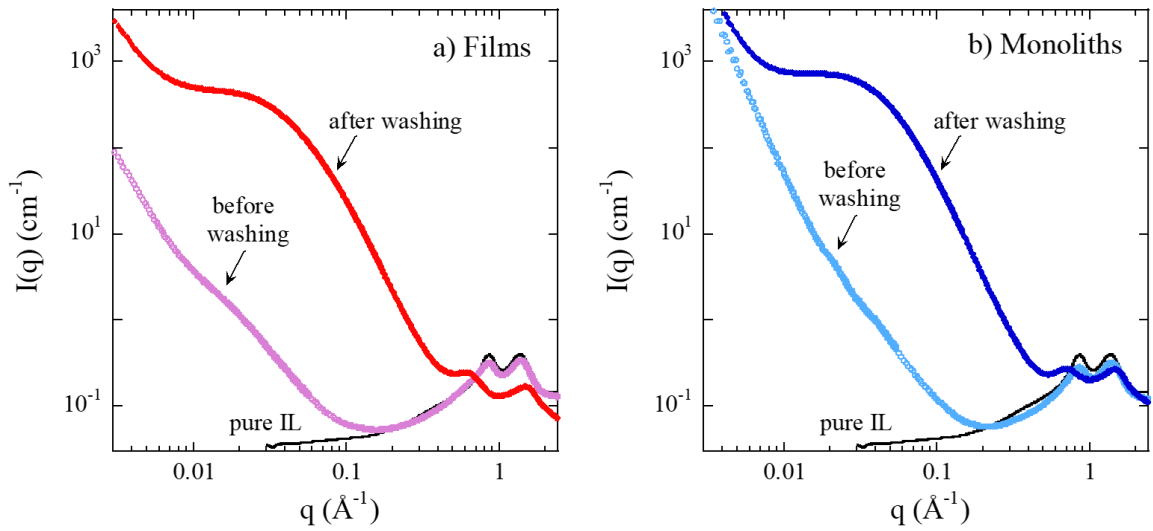

**Figure 5**: Small-angle X-ray scattering of (a) a film with 6 ml IL ($\Phi_{IL}$ = 86.3 %v) and (b) a monolith with 12 ml ($\Phi_{IL}$ = 80.8 %v) templating IL. Both are shown before and after washing, and the intensity of the pure ionic liquid (black line) is superimposed for comparison.

Once the film and monolith samples have been emptied, one can study their mesostructure by SAXS as a function of the amount of ionic liquid used for templating. In Figure 6, the resulting intensity curves in reduced presentation $I(q)/\Phi_{IS}$ for films and monoliths are shown. The curves overlap in the high-q range because of the intensity normalization procedure outlined in the methods section to obtain absolute units. Then, they separate into different intermediate-q plateaus. From visual inspection, it is found that these plateaus are higher in the case of the films, and that the corresponding Guinier domain is shifted to smaller angles. Both pieces of information hint towards a bigger nanostructure in the case of films, where the shift to the left hides a potential upturn which could exist (as with monoliths) at lower q but was not observed in the accessible q-window, down to $10^{-3}$ Å$^{-1}$. For monoliths, the low-q upturn is compatible with a $q^{-4}$ power law possibly generated by the macropore surface. From the ratio of the Porod prefactors at high and low q, the specific surface of the macropores is about 1000 times smaller than the one of the mesopores leading to an estimation of 1000 times bigger macropore sizes, thus typically about 2000 nm. This confirms the assumption of eq.(5) that the macropores do not contribute to the SAXS other than with the tail of their Porod domains. In case of the films, the ratio can be estimated only in rare cases with a beginning of a low-q upturn, and it is even bigger, of the order of $10^4$, which is why the low-q Porod domain is essentially unobserved. The order of the intermediate-q plateaus (shown in Figure 6 but in particular after normalization of their height by $\Phi_{meso}$ in Figures S9 and S10) follows the incorporated quantity of ionic liquid, higher $\Phi_{IL}$ giving a higher volume of the average mesopore. Figure 6 shows a representative set of all available samples with three 6 ml films and monoliths to highlight sample variability.

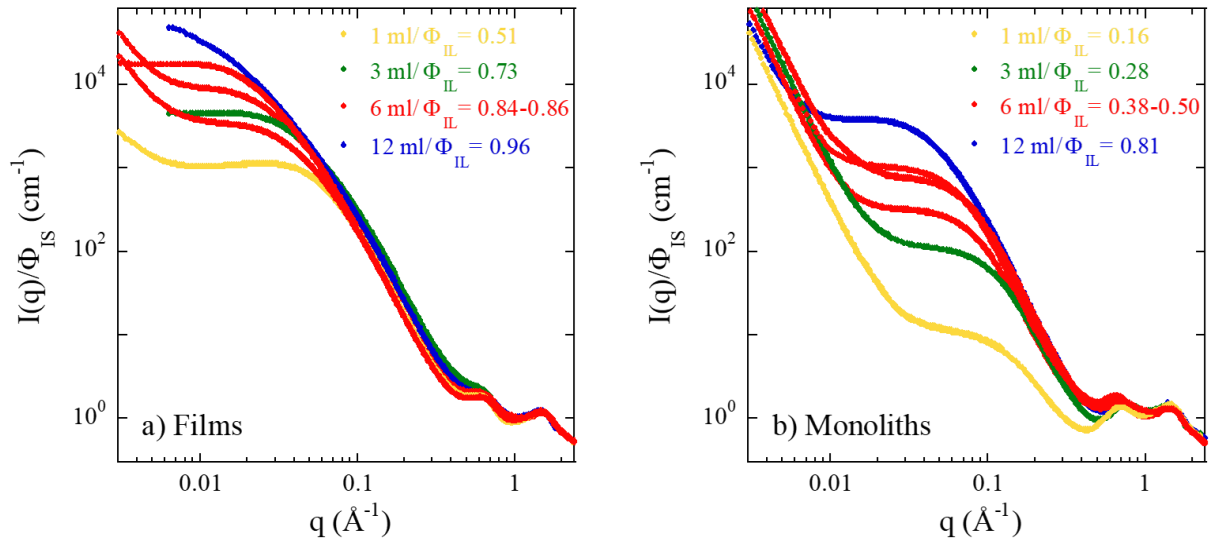

**Figure 6:** Scattered intensities for different IL-contents of emptied (a) films and (b) monoliths. The color code gives the quantity of templating-IL introduced as indicated by the legend. The corresponding IL volume fraction is also indicated.

The experimental scattering of the washed samples shown in Figure 6 can be compared to various geometric models (see Figure S11). In particular, only globular (or spherical) pores can be shown to be incompatible with the data, because of the too broad Guinier decay of the data, and because of the strong constraints connecting the total pore volume known from BET to the scattered intensity in absolute units. The key features of the intensity, namely the low-q intensity plateau and the position of the extended Guinier domain indicating the characteristic sizes, hint towards dominantly cylindrical pores of short length. Such a wormlike geometry has been proposed in the literature. [14,17,46] In this view, the exact shape of the curve in this range depends on the average cylinder dimensions, which can thus be determined by model fitting.

In Figure 7a, an exemplary fit of the SAXS data of a film is shown. First, if one started with cylinders of infinite length, i.e., in practice much longer than wide, one would encounter a well-known scattering signature in the low-q range scaling as $1/q$. The absence of such a low-q power law together with the presence of an intermediate scattering domain, show that cylinders of finite size, only a factor of 3 to 5 longer than the lateral radius, are present. Detailed modelling has been performed based on eq.(13) starting with only cylinders ($\beta = 1$). In presence of monodisperse cylinders in radius, the shape of the decay just below 0.1 Å$^{-1}$ would display strong oscillations. As this is not the case, and as we know from the BJH analysis that the pore width distribution is not monodisperse, we propose here an original combination of the BJH and the SAXS analysis. The preceding results obtained by TGA and adsorption isotherms provided the total mesopore volume fraction, i.e., the total amount of contrast-carrying nano-objects in scattering. Following eq.(9), the cylinder radius distribution function can be deduced from BJH, and used to determine the scattering contribution of polydisperse cylinders via eq.(7). By construction, the total pore volume is conserved using $\beta = 1$. The pure "cylinder"-function shown in Figure 7a corresponds to such a fit with a single fit parameter, the cylinder length, using the polydispersity in radius $N_{cyl}(R)$ deduced from the BJH distribution and shown in Figure 7b. This radial polydispersity is normed such that the area under the curve represents the total mesopore volume fraction of cylinders, i.e., we plot $N_{cyl} V_{cyl}(R) \, dR$. Based on the correct choice of $L_{cyl}$ – here we found $L_{cyl}$ = 12 nm –, the cylinder model is seen to reproduce the curvature of the Guinier domain and the following intensity decrease in Figure 7a ("cylinders only").

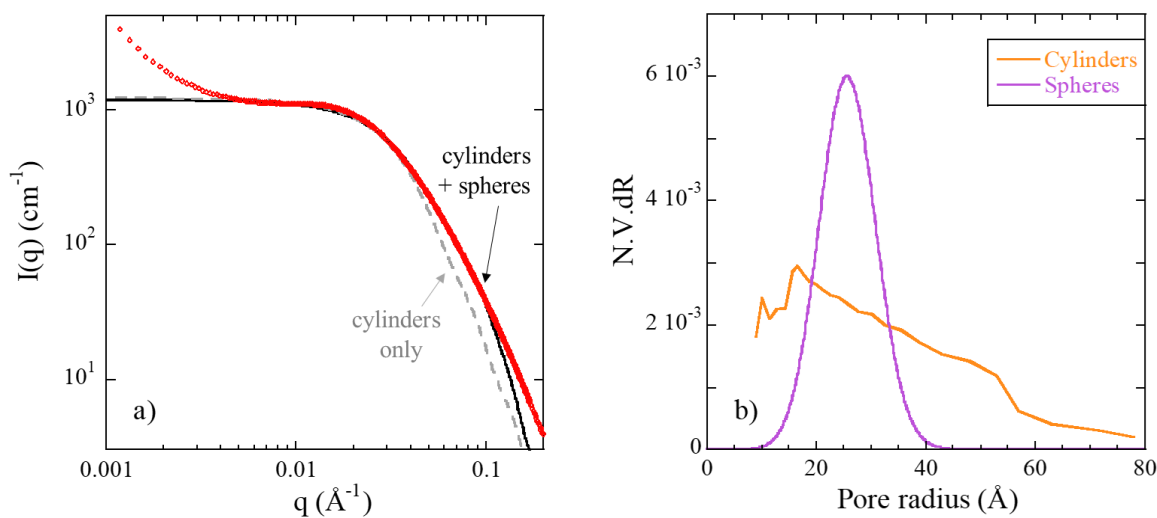

**Figure 7**: (a) Example of fit (black solid line) of emptied film sample (6 ml, symbols) by a superposition of cylinders of finite length and polydisperse in radius ($L_{cyl}$ = 12 nm, $<R_{cyl}>$ = 2.5 nm, β = 0.58) and polydisperse spheres ($<R_s>$ = 2.2 nm, σ = 25%, 1-β = 0.42). The pure cylinder contribution (β = 1) is included for comparison (dashed grey line). (b) Pore size distribution of cylinders calculated using eq.(9) from BJH data, and of spheres, expressed as contribution to the total volume fraction.

The radially polydisperse but otherwise perfectly straight "cylinders only" fit does not describe the data well above a critical wave vector of about 0.04 Å$^{-1}$. There must be an additional feature, most likely in the form of local deviations from cylindrical shape. Therefore, we have introduced an additive description of small globular pores following eq. (10). In this case, the cylinder contribution β is less than 1, in order for the mesopore volume to remain conserved. This additional contribution successfully prolongates the agreement of the theoretical scattering curve with the experimental one towards higher q, as shown by the full model including cylinders and spheres in Figure 6a. The mean size and polydispersity of the spheres need to be adapted to this q-range, and the best fit was obtained with the pore size distribution shown in Figure 7b, expressed again as the contribution to volume fraction, $N_s$ $V_s$ dR. It corresponds to $<R_s>$ = 2.2 nm and a polydispersity σ = 25% using a Gaussian distribution. It should be added that these small pores must be accessible, otherwise they could not be washed, and would thus either be visible in unwashed samples (if filled with air), or not visible in washed ones (if filled with IL). The contributions to the total pore volume fractions in Figure 7b show that in this case the cylinders and the spheres contribute about equally to the porosity, but due to the larger cylinder volume the latter dominate the scattering. In general, cylinders also dominate the pore volume fraction and can even make up 100% of the pore volume as shown in Figure S11.

As a cross-check, one may note that this model including the high-q treatment with spheres is based on the outcome of the BET and TGA results, which provide the radial polydispersity and $\Phi_{meso}$. Besides the geometrical parameters of average cylinder length and sphere radius (and their proportion β), all given by the shape of the intensity curve, there is only one free prefactor to the scattered intensity. The prediction coincides with the measured intensity within a factor 1.3 on average over all film and monolith samples (0.96 for the cylinder plus sphere model in Figure 7a; 0.72 for cylinders only). The minor discrepancies may be caused by imperfect knowledge of the scattering contrast.

The detailed fitting procedure has been applied to all film and monolith samples, with examples shown in Figures S12 and S13. It should be noted that it is not necessary to add spheres to the cylinders to obtain a fit of good quality in the high-q region for monoliths. Presumably, the cylinder length is sufficiently short approaching globular pores making superfluous an additional sphere contribution.

Besides, in the case of a low IL incorporation (1 ml and 3 ml monoliths), the samples have no or negligible mesoporosity (Figures 2 and 3) although the SAXS intensity in Figure 5 shows a pronounced Guinier regime which can be well described by small and very thin cylinders with a radial polydispersity of 25% (with $<R_{cyl}>$ = 0.5 – 1 nm, i.e., below the size of the other samples with BET signal, see Figure S14). It is not clear how these samples can have visible (and thus washable) pores, and we presume that the low mesoporosity is related to slow $N_2$ adsorption kinetics and difficult pore access on the time scale of the adsorption experiment, contrary to washing.

The model based on cylinders and spheres has been used to successfully fit all experimental data. The question of the relative importance of the two morphologies arises naturally, and we have plotted the corresponding (cylinder, respectively sphere) volume fractions in Figure S15, for both monoliths and films. The cylinders are seen to dominate the mesopore volume fraction. At highest IL contents, both contributions vanish as the samples then contain mostly macropores.

The main outcome of this study is summarized in Figure 8. We have regrouped all the dimensions of the cylinders and spheres. The cylinder length and radius, as well as the radii of the small spheres, are plotted as a function of $\Phi_{IL}$ for both films and monoliths. We underline that this representation automatically integrates the BET and TGA results, which have demonstrated that the films have higher porosity than the monoliths, and therefore their results extend to higher $\Phi_{IL}$-values in Figure 8. The striking feature of this figure is that all pore dimensions are found to overlap: the pore sizes are governed only by the real IL content. This means that the geometric dimensions of the pores – cylinders or spheres – are independent of the synthesis pathway, once syneresis is taken into account via the true $\Phi_{IL}$. They are found to increase in size as the IL-fraction increases, and notably above $\Phi_{IL}$ = 80% which is accessible only for films. At the highest IL-fractions, finally, the cylinder length is notably larger.

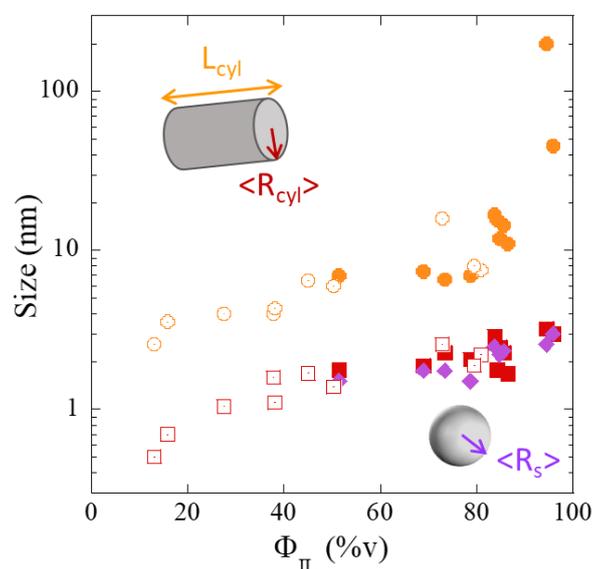

**Figure 8**: Summary plot of the mesopore geometries obtained from the SAXS analysis, as a function of the total incorporated IL volume fraction. All solid symbols refer to films, and all empty ones to monoliths. The orange discs give the cylinder length, the red squares the cylinder radii, and the purple diamonds the sphere radii. Spheres are not necessary for monoliths which are described only by short cylinders.

It is also interesting to note that the average cylindrical pore radius $<R_{cyl}>$ is very close to the one of the spheres $<R_s>$, and it also increases with $\Phi_{IL}$. We think that this means that all pores have roughly the same lateral size, and are more or less elongated. One could also imagine short cylinders connected by spherical pores of the same lateral dimensions. Finally, for the lowest $\Phi_{IL}$-fractions, the cylinder length approaches the radius, and these pores have almost globular shapes justifying the absence of spheres in the SAXS analysis. Upon addition of ionic liquid, there seems to be a tendency for both spheres and cylinders to grow with the IL content in Figure 8, although there is some scattering of the data. Additional IL thus widens the mesopores, while simultaneously increasing the quantity of macropores.

**Conclusion**

We have studied and compared the mesoporosity created by ionic liquid templating in two types of ionosilica samples, hydrolytic film samples of well-defined macroscopic dimensions, and monoliths as investigated by some of us before. [30-32] Thermogravimetric analysis has shown that the films incorporate the ionic liquid to a considerably higher extent. After washing out the ionic liquid, this part of the sample is converted into porosity. Nitrogen sorption has allowed to detect the part of mesoporosity, the rest corresponding to at least 1000 times bigger macropores, which are out of range of the BET analysis.

The main experimental parameter is the quantity of introduced templating ionic liquid. Adding more ionic liquid leads essentially to a higher formation of macropores for both types of samples. The highest fraction of mesoporosity is thus found for films with low ionic liquid contents. Following this result, a natural description of the samples in terms of the incorporated volume fraction of ionic liquid has been proposed. It allows presenting the SAXS analysis of the mesostructure in a unified way for all film and monolith samples. We have developed a conceptually new approach for data fitting, using the BJH analysis to describe the radial polydispersity of the cylinders. It is noteworthy that the prediction of the geometrical model based essentially on the fitted cylinder length, the measured BET mesoporosity, and the known contrast is very close to the experimentally observed intensity. In this context, we emphasize again the use of absolute units and a robust normalization procedure. As a result, a consistent set of fits and geometric pore parameters could be deduced for all samples after washing.

The general shape of the SAXS curves of emptied composites was discussed first. The intermediate-q scattered intensities are higher for the group of films, and the Guinier regime is more to the left, both indications hinting at larger mesopores in films than in monoliths. Our structural analysis of the SAXS data in absolute units in terms of a geometric model reveals the presence of both relatively short cylinders ($L_{cyl} \approx$ 5-10 nm, for radii ≈ 2 nm), and of small spheres (radius also 2 nm), for most samples. Only at very high IL incorporation (above 90%), longer cylinders are formed.

If one looks at the same samples not as a function of the introduced ionic liquid, but of the incorporated IL volume fraction, then the geometrical parameters are found to be superimposed on a master curve, regardless of the preparation pathway. The films with the higher incorporation efficiency are simply found on the right-hand-side of the plot. The addition of ionic liquid leads to higher mesopore volume fractions as seen by TGA and BET, and the corresponding pores are seen to grow.

The present detailed investigation of multi-scale porosity as a function of synthesis protocol offers exciting perspectives for future studies. First, in the context of creating new materials, one could now imagine refilling the emptied samples with solvent of specific properties, e.g., other ionic liquids for optimized ionic transport, or for further studies of the structure making use of contrast-matching opportunities for SANS experiments. Next, the precise description of the mesopores opens the door to confinement studies, for example of ionic liquids or polymer molecules, including poly(ionic liquid)s. The phase transition of confined ionic liquids, or the conformation of poly(ionic liquid)s in a mesoporous system of well-known geometry, are of fundamental interest for understanding ionic conductivity in view of applications in batteries or membranes.

**Acknowledgements.** This work was funded by the ANR IONOPILS project, Grant ANR-20-CE06-0027-01 of the French Agence Nationale de la Recherche. The authors acknowledge financial support from GDR2019 CNRS/INRAE "Solliciter LA Matière Molle" (SLAMM) and from the Synchrotron Soleil (beamline SWING). The authors also acknowledge E. Oliviero and V. Viguier from the MEA platform, Université de Montpellier, for the TEM experiments and sample preparation, respectively.

**Electronic supplementary information (ESI) available.** See DOI: XXX